\documentclass[journal]{IEEEtran}
    \usepackage{graphicx}
    \usepackage[numbers]{natbib}
    \usepackage{amsmath}
    \usepackage{amssymb}
    \usepackage[top=2cm, bottom=2cm, left=2cm, right=2cm]{geometry}
    \usepackage{algorithm}
    \usepackage{algorithmicx}
    \usepackage{algpseudocode}
    \usepackage{mathrsfs}
    \usepackage{calc}
    \usepackage{enumerate}
    \usepackage{indentfirst}
    \usepackage{graphicx, subfig}
	\allowdisplaybreaks[4]
    \title{Blockchain Aided Privacy-Preserving Outsourcing Algorithms of Bilinear Pairings for Internet of Things Devices}
    
\author{Hanlin~Zhang~\IEEEmembership{Member,~IEEE}, Le~Tong, Jia~Yu~\IEEEmembership{Member,~IEEE}, Jie~Lin
\thanks{This research is supported by National Natural Science Foundation of China (61572267, 61402245), Joint Found of the National Natural Science Foundation of China (U1905211), the Open Project of the State Key Laboratory of Information Security, Institute of Information Engineering, Chinese Academy of Sciences(2019-MS-03). Corresponding Author: Jia Yu

Hanlin~Zhang, Le~Tong and Jia Yu are with the School of Computer Science and Technology, Qingdao University, Qingdao 266071, China (Email: hanlin@qdu.edu.cn).

Hanlin~Zhang is also with the business school, Qingdao University, Qingdao 266071, China.

Jie~lin is with the School of Electronic and Information Engineering, Xi’an Jiaotong University, Xi’an 710049, China (Email: jielin@mail.xjtu.edu.cn).

}}
\begin{document}
    \maketitle
   	\section*{ABSTRACT}
Bilinear pairing is a fundamental operation that is widely used in cryptographic algorithms (e.g., identity-based cryptographic algorithms) to secure IoT applications. Nonetheless, the time complexity of bilinear pairing is $O(n^3)$, making it a very time-consuming operation, especially for resource-constrained IoT devices. Secure outsourcing of bilinear pairing has been studied in recent years to enable computationally weak devices to securely outsource the bilinear pairing to untrustworthy cloud servers. However, the state-of-art algorithms often require to pre-compute and store some values, which results in storage burden for devices. In the Internet of Things, devices are generally with very limited storage capacity. Thus, the existing algorithms do not fit the IoT well. In this paper, we propose a secure outsourcing algorithm of bilinear pairings, which does not require pre-computations. In the proposed algorithm, the outsourcer side's efficiency is significantly improved compared with executing the original bilinear pairing operation. At the same time, the privacy of the input and output is ensured. Also, we apply the Ethereum blockchain in our outsourcing algorithm to enable fair payments, which ensures that the cloud server gets paid only when he correctly accomplished the outsourced work. The theoretical analysis and experimental results show that the proposed algorithm is efficient and secure.

\
\begin{IEEEkeywords}
 Bilinear Pairings; Cloud Computing; Blockchain; Internet of Things
 \end{IEEEkeywords}
 
   	\section{INTRODUCTION}
 
The Internet of Things plays an essential role in the new generation of information technology~\cite{lin2017survey, yu2017survey}. It is known as the third wave of the world's information industry after computers and the Internet. With the rise of 5G in recent years, its characteristics of high reliability, ultra-low latency  and large-scale machine communication have driven the rapid development of the Internet of Things. Public-key cryptographic algorithms are widely applied in the IoT to safeguard the connected devices and the network, while some of the algorithms often involve time-consuming operations that the resource-constrained IoT devices cannot afford. For example, the bilinear pairing is a complex operation that is often applied in public-key cryptographic algorithms, such as the identity-based encryption algorithms and the elliptic curve cryptography algorithms. Thus, how to enable computationally weak IoT devices to accomplish complex operation is of great importance. 

Over the past decade, the development of the cloud computing~\cite{armbrust2010view} has seen explosive growth. Cloud servers provide on-demand computing services on a pay-as-you-go basis~\cite{vaquero2008break}, which allows users to delegate computation tasks to it and free themselves from the heavy computation workload. Taking advantages of the cloud computing, the computationally weak IoT devices can naturally accomplish complex cryptographic algorithms by outsourcing them to the cloud server. Nonetheless, outsourcing computation tasks to cloud servers also faces security challenges~\cite{ren2012security, yu2015enabling, yu2016enabling, yu2017strong}. First, the cloud server might be curious about the outsourced data, while the input and the output of an outsourced cryptography algorithm often involve sensitive information that should not be leaked to the cloud. For example, in the identity-based encryption, the input of a bilinear pairing involves the information about the private key, which should be kept secret from the untrustworthy cloud server. Thus, the privacy of the input and the output should be ensured during the outsourcing process. Second, the cloud server might return invalid computation results intentionally or unintentionally. The invalid result may be caused by software bugs or hardware failures. The cloud server might also return random results without executing the real computation tasks to save computation resources. Thus, IoT devices need to be able to verify if the returned results are correct during the outsourcing process. Also, the verification process and the workload to obscure the input and output should not involve any complex operation. The time cost of the outsourcing process on the client side should be unquestionably less than that of performing the original computation task on its own.

Secure outsourcing algorithms for bilinear pairing have been studied, and many existing algorithms can ensure the privacy of the input and output. Nonetheless, most of the existing outsourcing algorithms require a significant amount of pre-computations that are used for encryption and verification purposes. The large amount of the pre-computation data will bring significant demand for storage of IoT devices, while in fact, IoT devices are generally equipped with very limited storage space. Thus, existing secure outsourcing algorithms for bilinear pairing cannot properly be applied in the Internet of Things. 

Besides, in outsourcing computing, cloud/edge computing service providers should get paid only when correctly completing the computation task. Due to the lack of trust between outsourcers and users, traditional payment methods are difficult to ensure fairness. If cloud service providers get paid by the user first, it cannot be ensured that the cloud server will correctly perform the computation task; on the contrary, if the cloud service provider performs the computation task and returns the result to users first, it cannot guarantee that users will pay as they agreed on. Existing solutions to this trust problem usually rely on third parties such as banks, which will bring additional overhead, and trust is built on the basis of third parties. Fortunately, the emergence of blockchain and smart contract technology has made it possible to solve this problem. Based on the blockchain technology, the value can be directly transferred between the two parties in the form of cryptocurrencies without the need for a third party, which will provide a strong guarantee for the fairness of computing outsourcing services.

To address the above issues, in this paper, we explore how to securely outsource the bilinear pairing to untrustworthy cloud servers in a way that the IoT device does not need to perform pre-computations. Our contributions are summarized as follows:

\begin{itemize}

\item We propose a secure outsourcing algorithm for the bilinear pairing that does not require pre-computations. In the proposed algorithm, the input/output privacy of the client can be ensured so that the cloud server cannot obtain the original input/output.

\item We leverage the secure outsourcing of scalar multiplications in our proposed algorithm so that our algorithm does not require any pre-computations, and therefore the client does not need additional storage space to save pre-computation results. 

\item We develop a fair payment scheme which ensures that the cloud server can get paid only when he has correctly performed the outsourced computation task. We develop fair payment smart contracts on the Ethereum blockchain. 

\end{itemize}

This paper is an extension of our previous work~\cite{TYZ19}, which was published on 2019 IEEE Conference on Dependable and Secure Computing (DSC). Compared with the conference version, in this paper, we present the comparison of our proposed algorithm and state-of-art algorithms, both theoretically and experimentally. Moreover, we enhance the proposed algorithm by applying the Ethereum blockchain to enable fair payments.

The rest of the paper is organized as follows: We introduce some background knowledge, including bilinear pairings and elliptic curves in section~\ref{pre}. In section~\ref{mod}, we present the system model and provide some security definitions. We illustrate our developed secure outsourcing algorithms for bilinear pairings in section~\ref{alg}. In section~\ref{sec}, we analyze the security, efficiency and verifiability of the proposed algorithms. In section~\ref{block}, we propose a blockchain-based fair payment method which ensures that the server gets paid only when he correctly performed the outsourced computation task. In section~\ref{com}, we compare our proposed algorithms with state-of-art algorithms. We conduct both theoretical analysis and experiments to evaluate the performance of the proposed algorithms. In Section~\ref{rel}, we review the related works. Finally, we conclude the paper in Section~\ref{con}.

\section{Preliminaries}\label{pre}

In this section, we provide some background knowledge, including bilinear pairings, elliptic curves, and basic operations on elliptic curves. 

\subsection{Bilinear Pairings}

Bilinear pairing is a complex operation which is widely applied in cartographic protocols (e.g., one-round three-party key agreement, identity-based encryption). 
   	    	%$e:\ \mathbb{G}_1\ \times\ \mathbb{G}_2\ \to\ \mathbb{G}_T$
 
$\mathbb{G}_1$ and $\mathbb{G}_2$ are two cyclic additive groups, in which $P_1$ and $P_2$ are generators of $\mathbb{G}_1$ and $\mathbb{G}_2$, respectively. $p$, a large prime number, is the order of $\mathbb{G}_1$ and $\mathbb{G}_2$. Let $\mathbb{G}_T$ be a cyclic multiplicative group with the same order $p$. A bilinear pairing is a map $e\ :\ \mathbb{G}_1 \times \mathbb{G}_2\ \to\ \mathbb{G}_T$ which satisfies the following properties:
	   	    \begin{quote}
	   	    	\begin{enumerate}[(1)]
	   	    		\item Bilinear: $e(aR,\ bQ)\ =\ e(R,\ Q)^{ab}$ for any $R\in \mathbb{G}_1,\ Q\in \mathbb{G}_2$ and $a,\ b\in \mathbb{Z}^*_p$.
	   	    		\item Non-degenerate: There are $R\in \mathbb{G}_1$ and $Q\in \mathbb{G}_2$ such that $e(R,\ Q)\not=1$.
	   	    		\item Computable: There is an efficient algorithm to compute $e(R,\ Q)$ for all $R \in \mathbb{G}_1$ and $Q \in \mathbb{G}_2$.
	   	    	\end{enumerate}
	   	    \end{quote}
	   	    
\subsection{Elliptic Curve}
Suitable bilinear pairings can be constructed from the pairing for specially chosen elliptic curves. Assume $E$ is an elliptic curve defined on a finite field $\mathbb{F}_{p}$. The elliptic curve can be described as follows:
	   	    \begin{align}
	   	    	y^{2} = x^{3}+bx+c, \qquad \qquad b,c \in \mathbb{F}_{p}
	   	    \end{align}
	   	    The set of points $E(\mathbb{F}_{p})$ is a finite abelian group. We use $E=\{b,\ c,\ p\}$ to denote an elliptic curve. A point in $E(\mathbb{F}_{p})$ can be written as $(x,\ y)$.
	   	    
\subsection{Basic Operations on Elliptic Curve}

   	    \begin{enumerate}[(1)]
   	    	\item Point Addition: Let $P$ and $Q$ be two points on the elliptic curve, point addition describes the addition of $P$ and $Q$, which is denoted as point\_add($P$, $Q$) on elliptic curve. The point\_add($P$, $Q$) function first draws a straight line between the point $P$ and $Q$. The line intersects the elliptic curve at another point $-R$. The output of point\_add($P$, $Q$) is the point $R$, which is the reflection of the point $-R$ with respect to the X-axis.
   	    	
   	    	\item Point Doubling: Let $P$ be a point on the elliptic curve, point doubling describes the double of the point $P$, which is denoted as point\_double($P$). The function point\_double($P$) draws one tangent line to the elliptic curve at the point $P$. The line intersects the elliptic curve at the point $-R$. The output of point\_double($P$) is the point $R$, which is the reflection of the point $-R$ with respect to the X-axis.
   	    	
   	    	\item Scalar Multiplication: Let $P$ be a point on the elliptic curve, scalar Multiplication describes the operation that $P$ is multiplied by a scalar $n$, which is denoted as scalar\_multi($P$, $n$). The function scalar\_multi($P$, $n$) repeatedly adds the point $P$ to the result, which is denoted as $nP = P + P + P + ... + P$. 
   	    	
   	    	\item Double-and-add is the widely used approach to conduct the scalar multiplication operation on an elliptic curve. Let $P$ be a point on the elliptic curve, and let $n$ be a scalar. The binary form of scalar $n$ can be expressed as: $n_{m}n_{m-1}n_{m-2}...n_{0}$, where $m$ is the binary bits of $n$. The output is $Q = nP$. This algorithm works as follows:
   	    	\begin{align*}
	 	    	&Point\ Q\\
	 	    	&for\ i\ = \ 0\ to\ m\ do\\
	 	    	&\qquad if\ n_{i}\ =\ 1\ then\\
	 	    	&\qquad\qquad Q\ \leftarrow\ point\_add(Q,\ P)\\
	 	    	&\qquad end\ if\\
	 	    	&\qquad P\ \leftarrow\ point\_double(P)\\
	 	    	&end\ for\\
	 	    	&return\ Q
   	    	\end{align*}
   	    \end{enumerate}
	
	\section{System model and Security Definitions}\label{mod}

   		\subsection{System model}
   		
\begin{figure}[t]
		\centering
		\includegraphics[height=6.8cm,width=9.0cm]{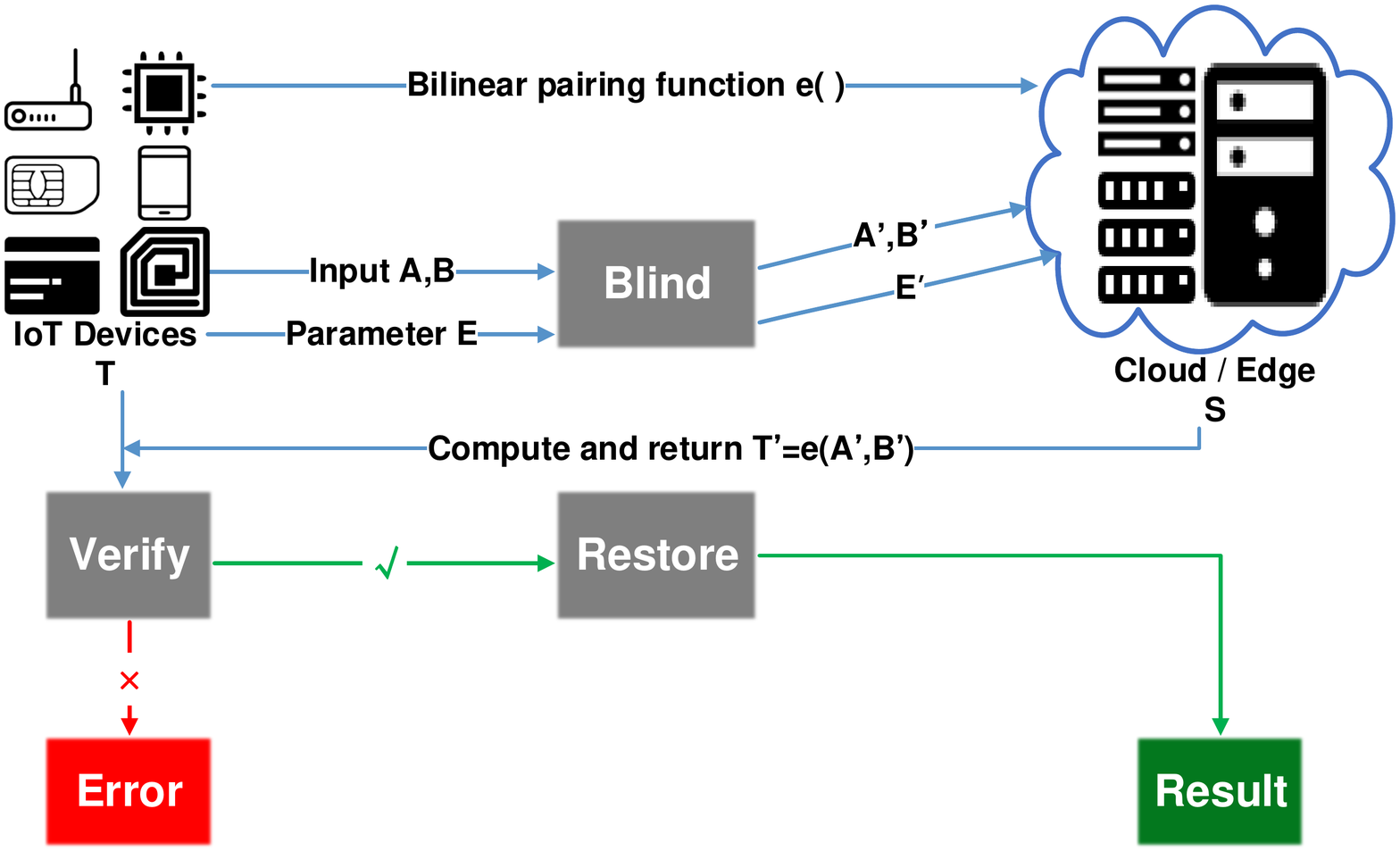}%height6.2 width9.0
		\caption{System model}
		\label{img_sysmodel}
	\end{figure}

	   	 Figure \ref{img_sysmodel} shows the workflow of outsourcing bilinear pairings. As shown in the figure, the IoT device $T$ and the cloud or edge server $U$ are the two parties involved in this framework. The IoT device $T$ needs to perform the complex bilinear pairing operation. However, $T$ is limited in computation power that he cannot afford such a time-consuming operation. The cloud or edge server $U$ is an entity with sufficient computation power that provides computation services at costs. Thus, $T$ plans to delegate the complex bilinear pairing to $U$. Before outsourcing the inputs ($A$ and $B$), $T$ transforms $A$ into $A'$ and $B$ into $B'$ respectively, to protect the privacy of the input and output. The server $U$ conducts the bilinear pairing operation with the obscured inputs ($A'$ and $B'$) and returns $e(A',B')$ to $T$. $T$ verifies the correctness of the returned result before recovering the original result. 
	   		   	    
   		\subsection{Security Definitions}
   		
   		We now introduce the security model for secure outsourcing of a cryptographic algorithm, which was proposed in \cite{hohenberger2005securely}. We use $Alg$ to denote the outsourced computation task. This security model contains three components: a resource constrained client $T$, an untrusted server $U$ and an untrusted environment $E$. $U$ and $E$ together play the role of the adversary $\mathcal{A}$, in which the environment $E$ writes programs for the server $U$ and submits adversarial chosen inputs to $Alg$. The resource constrained client $T$ outsources complex computations to $U$. $E$ and $U$ can communicate with each other only through $T$ once the program developed by $E$ is deployed on the client $T$. Suppose that instead of $U$, $T$ is given oracle access to a malicious $U'$. The adversary $\mathcal{A}=\{E,U^{'}\}$ tries to learn some information about the computation task $Alg$, including the input and output. Therefore, the goal of the client $T$ is to complete the computation task $Alg$ with the help from $U$. Informally, we say that $T$ securely outsources the computation task $Alg$ to the cloud server $U$ if: (a) $T$ and $U$ correctly  implement $Alg$, i.e., $Alg = T^{U}$. (b) even if the $U'$ is given oracle access to all the inputs and outputs of previous computation tasks, $U'$ still not able to obtain the input and output of $Alg$. Formally, we define an algorithm with outsource-I/O as follows:

	   		\textbf{Definition 1:(Algorithm with outsource-I/O)}.

Based on the level of secrecy, we categorize the inputs and outputs into 3 categories: (a) Secret: The information is only known to $T$, and is kept secret from $U$ and $E$. (b) Protected: The information is known to $T$ and $E$, and is kept secret from $U$. (c) Unprotected: The information is known to $T, U$ and $E$. Besides, based on whether the input is honest, we categorize the inputs into (a) honest and (b) adversarial. 
	   		
	   		Then, based on above categorizations, we divide inputs and outputs of $Alg$ as follows:
	   		
	   		\begin{quote}
	   			\textbf{Five types of inputs:}
	   			\begin{enumerate}[1)]
	   				\item The honest, secret input: The input that is generated by $T$, and is unknown to both $E$ and $U'$.
	   				\item The honest, protected input: The input that is generated by $T$, which is known to the environment $E$, but is protected from $U'$.
	   				\item The honest, unprotected input: The input that is generated by $T$, and is known to both $E$ and $U'$.
	   				\item The adversarial, protected input: The input that is generated by $E$, which is known to $T$, but protected from $U'$.
	   				\item The adversarial, unprotected input: The input that is generated by $E$, which is known by $T$ and $U'$.
	   			\end{enumerate}
	   			\textbf{Three types of outputs:}
	   			\begin{enumerate}[1)]
	   				\item The secret output: The output that is only known to $T$, but is unknown to both $E$ and $U'$.
	   				\item The protected output: The output that is known to $T, E$, but is unknown to $U'$.
	   				\item The unprotected output: The output that is known to $T, E$ and $U'$.
	   			\end{enumerate}
	   		\end{quote}

	   		\textbf{Definition 2:(Outsource-security).} Assume that an algorithm $Alg$ with above five types of inputs and three types of outputs. ($T,U$) is an outsourcing implementation of $Alg$. We say that ($T,U$) is outsource-secure if it meets the following requirements:
	   		\begin{quote}
	   			\begin{enumerate}[1)]
	   				\item Correctness: $T^{U}$ correctly implements $Alg$.
	   				\item Security: Assume $S_1$ and $S_2$ are polynomial-time simulators. The following pairs of random variables are computationally indistinguishable. 
	   			\end{enumerate}
   			\end{quote}
   			\textbf{Pair One:} $EVIEW_{real} \sim EVIEW_{ideal}() $:
   			\\
  			\indent $EVIEW_{real}$ is $E's$ view at the end of the following $REAL$ process.
   			\begin{flalign*}
	   			\begin{split}
	   				&EVIEW^{i}_{real}=(istate^{i},x^{i}_{hs},x^{i}_{hp},x^{i}_{hu}) \leftarrow \\
   					&I(1^{k},istate^{i-1});\\
   					&(estate^{i},j^{i},x^{i}_{ap},x^{i}_{au},stop^{i}) \leftarrow\\ &E(1^{k},EVIEW^{i-1}_{real},x^{i}_{hp},x^{i}_{hu});\\
   					&(tstate^{i},ustate^{i},y^{i}_{s},y^{i}_{p},y^{i}_{u})\leftarrow\\ &T^{U^{'}(ustate^{i-1})}\\
   					&(tstate^{i-1},x^{j^{i}}_{hs},x^{j^{i}_{hp}},x^{j^{i}_{hu}},x^{i}_{ap},x^{i}_{au}):\\
   					&(estate^{i},y^{i}_{p},y^{i}_{u})
	   			\end{split}
   			\end{flalign*}
   			\[EVIEW_{real}=EVIEW^{i}_{real}\ if\ stop^{i}=TRUE.\]
The real process runs in rounds. In round $i$, an honest, stateful process $I$ first generates honest inputs, including honest, secret input $(x^{i}_{hs})$,  honest, protected input $x^{i}_{hp}$ and honest, unprotected input $x^{i}_{hu}$. Based on $x^{i}_{hp}$, $x^{i}_{hu}$ and its view from last round $EVIEW^{i-1}_{real}$, the environment $E$ then generates the following 5 outputs: 
   				\begin{quote}
   				  \begin{enumerate}[1)]
	   				\item $estate_{i}$: the internal state of current round. 
	   				\item $j^{i}$: specifies which previously generated honest input to be given to $T^{U^{'}}$. 
	   				\item $x^{i}_{ap}$: the adversarial, protected input.
	   				\item $x^{i}_{au}$: the adversarial, unprotected input. 
	   				\item $stop^{i}$: the boolean variable which decides whether to stop the process in the current round. 
	   			\end{enumerate}
	   			\end{quote}
	   			
Then the algorithm $T^{U^{'}}$ takes as input $T's$ state of last round $tstate^{i-1}$, $U's$ state of last round $ustate^{i-1}$ and previously generated 5 inputs $(x^{j^{i}}_{hs},x^{j^{i}}_{hp},x^{j^{i}}_{hu},x^{i}_{ap},x^{i}_{au})$. The algorithm outputs $T's$ state of the current round $tstate^{i}$, $U's$ state of the current round $ustate^{i}$, the secret output $y^{i}_{s}$, the protected output $y^{i}_{p}$ and unprotected output $y^{i}_{u}$. $(estate^{i},y^{i}_{p},y^{i}_{u})$ is the view of the real process in round $i$. $EVIEW_{real}$, the final view of $E$ in the real process is the view of the last round. 
\begin{figure*}[htb]
	\centering
	\includegraphics[width=16.0cm]{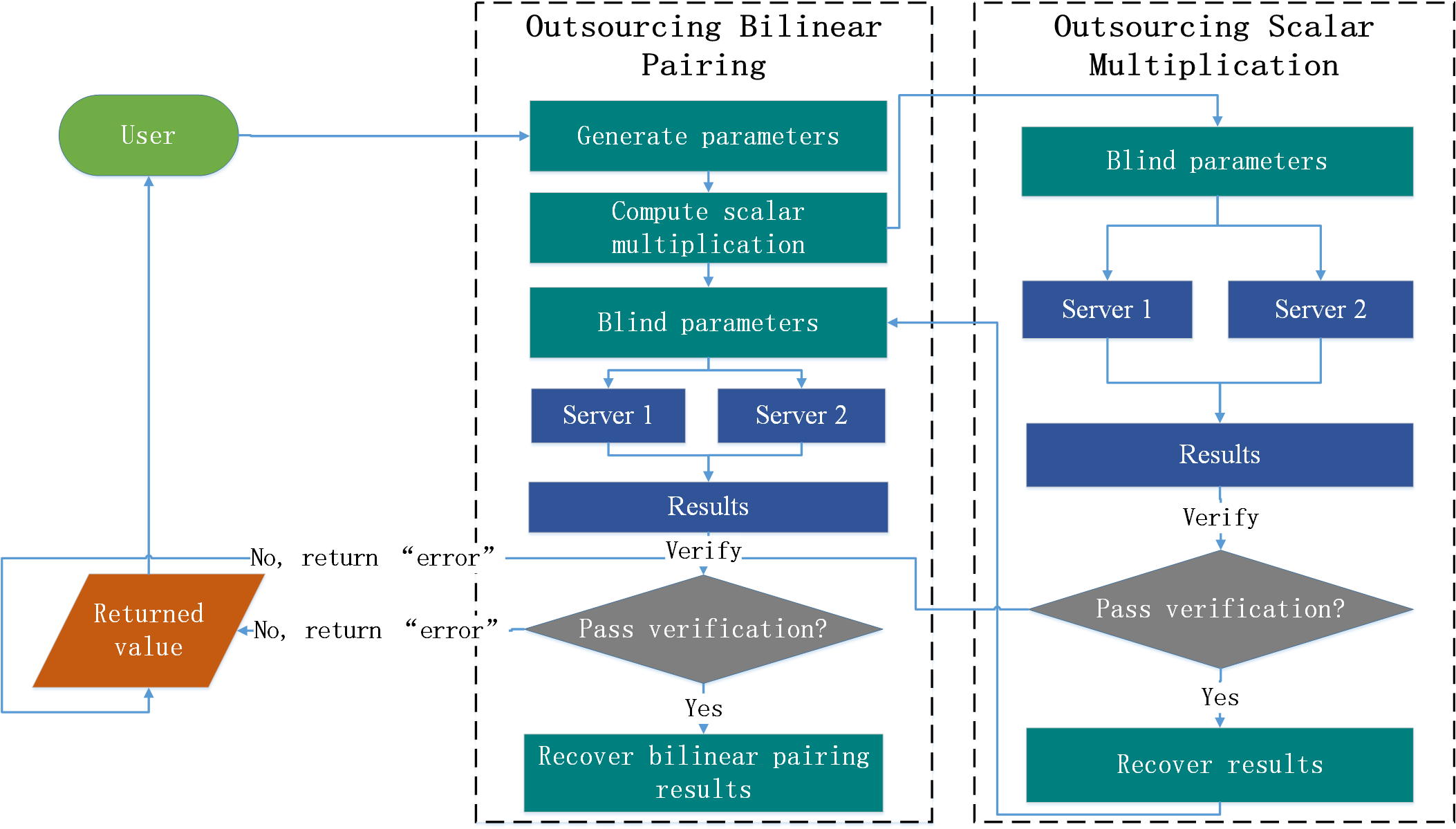}%height6.2 width9.0
	\caption{Outsourcing workflow}
	\label{oswork}
\end{figure*}

$EVIEW_{ideal}$ is $E's$ view at the end of the following $ideal$ process:
   			\begin{flalign*}
   			    \begin{split}
	   		        &EVIEW^{i}_{ideal}=(istate^{i},x^{i}_{hs},x^{i}_{hp},x^{i}_{hu}) \leftarrow \\
	   				&I(1^{k},istate^{i-1});\\
	   				&(estate^{i},j^{i},x^{i}_{ap},x^{i}_{au},stop^{i}) \leftarrow\\ &E(1^{k},EVIEW^{i-1}_{ideal},x^{i}_{hp},x^{i}_{hu});\\
	   				&(astate^{i},y^{i}_{s},y^{i}_{p},y^{i}_{u})\leftarrow\\
	   				&Alg(astate^{i-1},x^{j^{i}}_{hs},x^{j^{i}}_{hp},x^{j^{i}}_{hu},x^{i}_{ap},x^{i}_{au});\\
	   				&(sstate^{i},ustate^{i},Y^{i}_{p},Y^{i}_{u},replace^{i})\leftarrow\\
	   				&{S_1}^{U^{'}(ustate^{i-1})}(sstate^{i-1},...\\
	   				&...x^{j^{i}}_{hp},x^{j^{i}}_{hu},x^{i}_{ap},x^{i}_{au},y^{i}_{p},y^{i}_{u});\\
	   				&(z^{i}_{p},z^{i}_{u})=replace^{i}(Y^{i}_{p},y^{i}_{u})+\\
	   				&(1-replace^{i})(y^{i}_{p},y^{i}_{u}):\\
	   				&(estate^{i},z^{i}_{p},z^{i}_{u})
   			    \end{split}
   			\end{flalign*}
   			
   			\[EVIEW_{ideal}=EVIEW^{i}_{ideal}\ if\ stop^{i}=TRUE.\]
   			
 In this process, the stateful simulator $S_{1}$ simulates the view of $E$, who does not have the access to $x^i_hs$, but has the access to non-secret outputs of $Alg$. In round $i$, $S_1$ takes as input its state from last round $s_1state^{i-1}$, $U$'s state from last round $ustate^{i-1}$, non-secret inputs of $Alg$ $(x^{j^{i}}_{hp},x^{j^{i}}_{hu},x^{i}_{ap},x^{i}_{au})$, and non-secret outputs of $alg$ $(y^{i}_{p},y^{i}_{u})$, and outputs its state of the current round $s_1state^i$, $U$'s state of the current round $ustate^i$, fake output values $(Y^{i}_{p}$, $Y^{i}_{u})$, and an indicator variable $replace^{i}$. The indicator variable $replace^{i}$ indicates whether to replace the original outputs $(y^{i}_{p},y^{i}_{u})$ with fake outputs $(Y^{i}_{p}$, $Y^{i}_{u})$. I.e., if $replace^{i}$ equals 1, the final output $(z^{i}_{p}, z^{i}_{u})$ equals $(Y^{i}_{p}$, $Y^{i}_{u})$. If $replace^{i}$ equals 0, the final output $(z^{i}_{p}, z^{i}_{u})$ equals $(y^{i}_{p},y^{i}_{u})$. The final view of $E$ in the ideal process $EVIEW_{ideal}$ is $(estate^{i},z^{i}_{p},z^{i}_{u})$.

	   		\textbf{Pair Two:} $UVIEW_{real} \sim UVIEW_{ideal}() $:
	   		\\ 
The view that the untrusted software $U^{'}$ obtains by participating in the REAL process is described in \textbf{Pair One}. \[UVIEW_{real}=ustate^{i}\ if\ stop^{i}=TRUE.\]
   			The \textbf{IDEAL} process:
   			\begin{flalign*}
   			    \begin{split}
	   				&UVIEW^{i}_{ideal}=(istate^{i},x^{i}_{hs},x^{i}_{hp},x^{i}_{hu}) \leftarrow \\
	   				&I(1^{k},istate^{i-1});\\
	   				&(estate^{i},j^{i},x^{i}_{ap},x^{i}_{au},stop^{i}) \leftarrow\\ &E(1^{k},estate^{i-1},x^{i}_{hp},x^{i}_{hu},y^{i-1}_{p},y^{i-1}_{u});\\
	   				&(astate^{i},y^{i}_{s},y^{i}_{p},y^{i}_{u})\leftarrow\\
	   				&Alg(astate^{i-1},x^{j^{i}}_{hs},x^{j^{i}}_{hp},x^{j^{i}}_{hu},x^{i}_{ap},x^{i}_{au});\\
	   				&(sstate^{i},ustate^{i})\leftarrow\\
	   				&S^{U^{'}}_{2}(sstate^{i-1},x^{j^{i}}_{hu},x^{i}_{au}):\\
	   				&(ustate^{i})
   			    \end{split}
   			\end{flalign*}
   			\[SVIEW_{ideal}=SVIEW^{i}_{ideal}\ if\ stop^{i}=TRUE.\]
   			\indent The stateful simulator $S_{2}$ in the ideal process is similar with $S_1$ in Pair One. $S_{2}$ only takes the unprotected $(x^{i}_{hu},x^{i}_{au})$ to query $U^{'}$.

\textbf{Definition 3} ($\beta$-checkable \cite{hohenberger2005securely}): Checkability requires that the client $T$ could detect the invalid results from the cloud server $U$ with high probability. An algorithm \emph{(T,U)} is said to be $\beta$-checkable if (a) the client $T$ and the cloud server $U$  perform the algorithm correctly and (b) for $\forall$\emph{x}, if the server $U$ misbehaves during execution of  ${T^{U}}(x)$, \emph{C} could detect it with probability greater than or equals $\beta$.

\textbf{Definition 4} ($\alpha$-efficient \cite{hohenberger2005securely}): Efficiency requires that the workload carried by \emph{T} should be significantly less than conducting the original computation on its own. An algorithm \emph{(T,U)} is $\alpha$-efficient if (a) $T$ and $U$ correctly implement the algorithm and (b) for $\forall$\emph{x}, the execution time of ${T^U}$ is less than or equal to an $\alpha$-multiplicative factor of the execution time of \emph{F}.
	   			   		
\section{The Proposed Secure Outsourcing Algorithm}\label{alg}
\subsection{Design Rationale}	
We now describe our underlying thoughts when we design the proposed algorithm. The overall objective is to design an algorithm that can protect the input/output privacy and the checkability when outsourcing the bilinear pairing. Meanwhile, the proposed scheme should not require extra storage space. The existing secure outsourcing algorithms often require a significant amount of pre-computations, which would bring huge demand for storage space of IoT devices. Thus, when we design the outsourcing algorithm, we focus on developing a strategy that avoids the pre-computation. To obscure the input and output of bilinear pairings, we consider multiplying the input points $A$ and $B$ by random scalars. Notice that the scalar multiplication itself is a time-consuming operation. Thus, we consider outsourcing the scalar multiplication as well. Figure~\ref{oswork} shows the overall workflow of the proposed algorithm. The local resource-constrained IoT device obscures the scalar multiplication and sends it to the cloud server. The cloud server returns the result after conducting the scalar multiplication, and the user decides whether to accept it after verifying the correctness of the result. After recovering the result of the scalar multiplication, the user sends the calculated points to the cloud. On receiving the inputs, the cloud server calculates the bilinear pairing and returns the calculation result to the user. After the user receives the result, the correctness verification is performed. If the returned result passes the verification, the user recovers the original result from it.

\subsection{$SM$: The secure outsourcing algorithm for scalar multiplications}
As introduced above, when obscuring the inputs of the bilinear pairing, the client needs to outsource the scalar multiplications to the cloud server. The state-of-art algorithm to outsource the scalar multiplication was proposed by Zhou {\em et al.} in \cite{zhou2017expsos}. In their proposed scheme, they outsource two relevant scalar multiplications to a single cloud server. Since we have two cloud servers in our system model, we leverage their strategy but adjust the strategy to fit the one-malicious version of the two-untrusted-program model. We now introduce the algorithm to outsource the scalar multiplication in our system.

  \begin{algorithm}[H]
	   	        \caption{SM}
	   	        \label{alg:sm}
	   	        \begin{algorithmic}[1]
	   	            \Require
	   	            	$E =\{a, b, p\}$, $P(x, y) \in E(\mathbb{F}_{p})$ and $c \in \mathbb{F}_{p}$
	   	    		\Ensure
	   	    			$R\ =\ cP$
	   	    		\begin{enumerate}
	       			    \item \textbf{Problem Transformation}
	       			    \begin{enumerate}
	       			   		 \item
	   	    					The client generates a random prime $q$ and computes $N = pq$.
	       			    	\item
	       			    		The client selects random integers $r_1,r_2,r_3,r_4,r_5,r_6,t_1,t_2$ and calculates:
	       		
		       			    	\begin{flalign*}
			       			    	x'\ &=\ (x\ +\ r_1p) mod\ N&\\
			       			    	y'\ &=\ (y\ +\ r_2p) mod\ N&\\
			       			    	a'\ &=\ (a\ +\ r_3p) mod\ N&\\
			       			    	b'\ &=\ (b\ +\ r_4p) mod\ N&\\
			       			    	c_1\ &=\ (c\ +\ r_5p) mod\ N&\\
			       			    	c_2\ &=\ (t_1c\ +\ t_2\ +\ r_6p) mod\ N&\\
			       			    	P'\ &=\ (x',\ y')&
		       			    	\end{flalign*}
		       			    	
	       			    \end{enumerate}
	       		        \item \textbf{Computation}
	       		        \begin{enumerate}
	       		        	\item
	       		        		Client queries $U_1$ as:
	       		        		$Q_1\ =\ c_1P'$
	       		        		$Q_3\ =\ r_1Q_1\ +\ r_2P'$
	       		        	\item
	       		        		Client queries $U_2$ as:
	       		        		$Q_2\ =\ c_2P'$
	       		        	\item
	       		        		$U_1$ returns $Q_1$ and $Q_3$ to Client, $U_2$ returns $Q_2$ to Client.
	       		        \end{enumerate}
	       	            \item \textbf{Verification}
	       	            \begin{enumerate}
	       	            	\item
	       	            		Client verifies the results by verifying $Q_3\ \stackrel{?}{=}\ Q_2\ (mod\ p)$.
	       	            \end{enumerate}
	                    \item \textbf{Recovery}
	                    \begin{enumerate}
	                    	\item
	                    		Client recovers the result $R\ =\ Q_1\ (mod\ p)$.
	                    \end{enumerate}
	   	    		\end{enumerate}
	   	    		\label{code:fram:select}
	   	        \end{algorithmic}
	   	    \end{algorithm}
	
The algorithm is named $SM$, which is shown in algorithm~\ref{alg:sm}. The input of the algorithm includes the elliptic curve $E =\{ a, b, p\}$, a point on the curve $P(x, y) \in E(\mathbb{F}_{p})$, and a scalar $c \in \mathbb{F}_{p}$. The algorithm outputs a point $R = cP$ on the elliptic curve. The client first select a random prime number $q$, which is with the same length of $p$, and calculates $N = pq$. Then, the client selects random integers $r_1, r_2$ calculates $x' = (x + r_1p) \bmod N$, $y' = (y + r_2p) \bmod N$ to obscure the point $P$. The client selects random integers $r_3, r_4$ and calculates $a' = (a + r_3p) \bmod N$, $b' = (b + r_4p) \bmod N$ to obscure the parameters of the elliptic curve $E$. The client selects a random integer $k_5$ and calculates $c_1 = (c + r_5p) \bmod N$ to blind the scalar $c$. For verification purpose, the client also selects a random integer $k_6$ and calculates $c_2 = (t_1c + t_2 +\ r_6p) \bmod N$. Then the client queries $U_1$ and obtain $Q_1\ =\ c_1P'$, $Q_3\ =\ r_1Q_1\ +\ r_2P'$. The client queries $U_2$ and obtains $Q_2\ =\ c_2P'$. On receiving the returned results, the client verifies if $Q_3 =  Q_2 (mod\ p)$. If the results pass the verification, the client recovers original result $R\ =\ Q_1\ (mod\ p)$.

   	    \subsection{$BPSM$: Our proposed secure algorithm}
   	    
   	      \begin{algorithm}[H]
	   	    	\caption{BPSM}
	   	    	\label{alg:Framwork}
	   	    	\begin{algorithmic}[1]
	   	    		\Require
	   	    			$ A\in\mathbb{G}_1,\ B\in\mathbb{G}_2 $ and $E =\{a, b, p\}$. 
	   	    		\Ensure
	   	    			$e(A,\ B)$
	   	    		\begin{enumerate}
	   	    			\item
	   	    				Client randomly selects four integer $a_1$, $a_2$, $b_1$ and $b_2$, such that $a_1a_2+b_1b_2\ =\ 1$. Client selects an random integer $x$.
	   	    			\item
	   	    				Client runs $SM()$ to obtain:
	   	    			\begin{flalign*}
		   	    			a_1A\ &=\ SM(a_1,\ A)&\\
		   	    			b_1A\ &=\ SM(b_1,\ A)&\\
		   	    			a_2B\ &=\ SM(a_2,\ B)&\\
		   	    			b_2B\ &=\ SM(b_2,\ B)&\\
		   	    			xb_1A\ &=\ SM(x,b_1A)&\\
		   	    			xa_2B\ &=\ SM(x,a_2B)&
	   	    			\end{flalign*}
	   	    			\item
	   	    				Client queries $U_1$ in random order as:
	   	    			\begin{flalign*}
		   	    			H_1\ &=\ U_1(a_1A,\ a_2B)\ =\ e(A,B)^{a_1a_2}&\\
		   	    			L_1\ &=\ U_1(xb_1A,\ b_2B)\ =\ e(A,\ B)^{xb_1b_2}&
	   	    			\end{flalign*}
	   	    			Similarly, Client queries $U_2$ in random order as:
	   	    			\begin{flalign*}
		   	    			H_2\ &=\ U_2(b_1A,\ b_2B)\ =\ e(A,B)^{b_1b_2}&\\
		   	    			L_2\ &=\ U_2(xa_1A,\ a_2B)\ =\ e(A,\ B)^{xa_1a_2}&
	   	    			\end{flalign*}
	   	    			\item
	   	    			Finally, Client verifies the results by checking $L_{1}L_{2}\ \stackrel{?}{=}\ (H_{1}H_{2})^x$. If the equality does not hold, client outputs "error". Otherwise, because $a_1a_2\ +\ b_1b_2\ =\ 1$, client can compute $e(A,\ B)\ =\ H_{1}H_{2}$.
	   	    		\end{enumerate}
	   	    		\label{code:fram:select}
	   	    	\end{algorithmic}
	   	    \end{algorithm}

   	    We now introduce our proposed secure algorithm for bilinear pairings, which is named $BPSM$. $BPSM$ takes the elliptic curve $E$ and two points on the curve $A$ and $B$ as the input, and output $e(A, B)$. The algorithm runs as follows: The client first generates random integers $a_1$, $a_2$, $b_1, b_2$, s.t., $a_1a_2+b_1b_2\ =\ 1$. The client also generates a small integer $x$. With the generated parameters, the client calls the $SM$ function to calculate a set of scalar multiplications and obtains $a_1A = SM(a_1, A)$, $a_2A = SM(a_2, A)$, $b_1A = SM(b_1, A)$, $b_2A = SM(b_2, A)$, $xb_1A = SM(x, b_1 A)$, $xa_2A = SM(x, a_2A)$. $a_1A, a_2A, b_1A, b_2A, xb_1A$ and $xb_2A$ are now the obscured inputs to conduct the bilinear parings. The client queries $U_1$ in a random order and obtains $H_1 =\ U_1(a_1A,\ a_2B)\ =\ e(A,B)^{a_1a_2}$ and  $L_1\ =\ U_1(xb_1A,\ b_2B)\ =\ e(A,\ B)^{xb_1b_2}$. The client the queries $U_2$ in a random order and obtains $H_2\ =\ U_2(b_1A,\ b_2B)\ =\ e(A,B)^{b_1b_2}$ and $L_2\ =\ U_2(xa_1A,\ a_2B)\ =\ e(A,\ B)^{xa_1a_2}$. Based on the returned results, the client verifies if $L_{1}L_{2}\ = (H_{1}H_{2})^x$. If the results pass the verification, the client recovers original result $e(A,\ B)\ =\ H_{1}H_{2}$. Otherwise, the client outputs "error".

	    \section{Security Analysis}\label{sec}
		    \textbf{Theorem 1}. The algorithms ($T$,($U_{1},U_{2}$)) of \textbf{BPSM} are an outsource-secure implementation, where the input $(A,B)$ may be honest, secret, or honest, protected, or adversarial, protected.\\
		    \textbf{PROOF}. According to Definition 2, we need to prove the correctness and the security. We first prove the correctness:\\
		    \begin{flalign*}
		    \begin{split}
			    H_{1}H_{2}&=U_{1}(u_{1}A, u_{2}B)U_{2}(v_{1}A, v_{2}B)\\
			    &=e(A,B)^{u_{1}u_{2}}e(A,B)^{v_{1}v_{2}}\\
			    &=e(A,B)^{u_{1}u_{2}+v_{1}v_{2}}\\
			    &=e(A,B)
		    \end{split}
		    \end{flalign*}
		    \indent We now prove the security.
		    \begin{itemize}
		    	\item \textbf{Pair One:} $EVIEW_{real} \sim EVIEW_{ideal}$:\\
		    	\indent If the input ($A, B$) is not honest and secret, the performance of $S1$ is the same with that in the real execution, and $EVIEW_{ideal}$ is therefore same with $EVIEW_{real}$. Thus, we suppose that the input ($A, B$) is honest and secret. In round $i$, the stateful simulator $S_1$ behaves as follows. When receiving the inputs, $S_1$ ignores the inputs and randomly selects points ($w_{1}P_{1}, w_{2}P_{2}, w_{3}P_{1}, w_{4}P_{2}$) and ($w_{5}P_{1}, w_{4}P_{2}, w_{6}P_{1}, w_{2}P_{2}$) and makes random queries to $U'_{1}$ and $U'_{2}$ as follows:
		    	
		    	\begin{flalign*}
		    	\begin{split}
		    	&U_{1}(w_{1}P_{1}, w_{2}P_{2}) \rightarrow H_{1}\\
		    	&U_{1}(w_{3}P_{1}, w_{4}P_{2}) \rightarrow L_{1}\\
		    	&U_{2}(w_{5}P_{1}, w_{4}P_{2}) \rightarrow H_{2}\\
		    	&U_{2}(w_{6}P_{1}, w_{2}P_{2}) \rightarrow L_{2}
		    	\end{split}
		    	\end{flalign*}
		    	
		    	\indent Based on the the results from $U'_{1}$ and $U'_{2}$, $S_{1}$ performs as follows:
		    	\begin{enumerate}[-]
		    		\item If there exists an error, $S_1$ outputs $Y^{i}_{p}=$"ERROR", $Y^{i}_{u}=\emptyset$, $replace^{i}=1$. $S_1$ saves the state of the current round.
		    		\item If no error is detected, $S_{1}$ outputs $Y^{i}_{p}=\emptyset$, $Y^{i}_{u}=\emptyset$, $rep^{i}=0$. $S_1$ saves the state of the current round.
		    		
		    	\end{enumerate}
		    	\indent The inputs to ($U'_{1}$, $U'_{2}$) between the real process and ideal process are computationally indistinguishable. The reason is that in the ideal process, the inputs are chosen randomly by $S_1$. In the real process, the inputs of $U_{1}$ and $U_{2}$ are independently re-randomized. 
		    	\indent We have the following possible cases to be considered:
		    	\begin{enumerate}[-]
		    		\item $U_{1}$ and $U_{2}$ perform honestly in the round $i$. In this case, $T^{U^{'}_{1},U^{'}_{2}}$ correctly implements \textbf{BPSM} in the real experiment and $S_{1}$ will not replace the output in the ideal experiment, then $EVIEW_{real} \sim EVIEW_{ideal}$. 
		    		\item One of $U_{1}$ and $U_{2}$, or both of them behave dishonestly, the misbehavior will be detected by $S_{1}$ and $T$, the algorithm outputs "ERROR". In this process, $EVIEW_{real} \sim EVIEW_{ideal}$.
		    	\end{enumerate}
		    	
  Thus, we can conclude that whether $U_{1}$ and $U_{2}$ misbehave or not, $EVIEW_{real} \sim EVIEW_{ideal}$. 
		    		
		    	\item \textbf{Pair Two:} $UVIEW_{real} \sim UVIEW_{ideal}$:\\
		    	In round $i$, the stateful simulator $S_{2}$ performs similar with $S_{1}$. When receiving the input, $S_{2}$ ignores it and randomly selects points and makes random queries to $U_{1}$ and $U_{2}$. $S_{2}$ then saves the state of $U_1$ and $U_2$ and also its own state. The inputs generated by $T$ are randomized and independent. Thus, we can conclude that $UVIEW_{real} \sim UVIEW_{ideal}$
		    	
		    \end{itemize}
		    \textbf{Theorem 2}. The algorithms ($T$,($U_{1}$,$U_{2}$)) of \textbf{BPSM} is 1-checkable.
		    
		    \textbf{PROOF}. When receiving the results form the cloud servers, the client checks if the equation $L_{1}L_{2}=(H_{1}H_{2})^x$ holds. If $U_1$ and $U_2$ misbehave, The client will detect the misbehavior with probability 1. Thus, according to definition 3, the algorithms ($T$,($U_{1}$,$U_{2}$)) of \textbf{BPSM} is 1-checkable.
		    
		    \textbf{Theorem 3}. The algorithms ($T$,($U_{1}$,$U_{2}$)) of \textbf{BPSM} is $\frac{1}{logp}$-efficient.
		    
		    \textbf{PROOF}. The algorithm \textbf{BPSM} requires 1 modular exponentiation and 2 multiplication in $\mathbb{G}_T$. The original bilinear pairing requires roughly $log p$ multiplications in resulting finite field \cite{chen2015efficient}. Thus, according to definition 4, the algorithms ($T$,($U_{1}$,$U_{2}$)) of \textbf{BPSM} is $\frac{1}{logp}$-efficient.

    \section{Blockchain-based fair payment scheme}\label{block}
    
     In outsourcing computing, cloud/edge computing service providers should get paid only when correctly completing the computation task. Due to the lack of trust between service providers and users, traditional payment methods are difficult to ensure fairness. If cloud service providers get paid by the user first, it cannot be ensured that the cloud server will correctly perform the computation task; on the contrary, if the cloud service provider performs the computation task and returns the result to users first, it cannot guarantee that users will pay as they agreed on. Existing methods to solve this trust problem are based on traditional electronic cash, relying on third parties such as banks, which will bring additional overhead, and trust is built on the basis of third parties. Fortunately, the emergence of blockchain and smart contract technology has made it possible to solve this problem. Based on the blockchain technology, the value can be directly transferred between the two parties in the form of cryptocurrencies without the need for a third party, which will provide a strong guarantee for the fairness of computing outsourcing services.

\begin{figure*}
  	\centering
    	\includegraphics[width=6.0in]{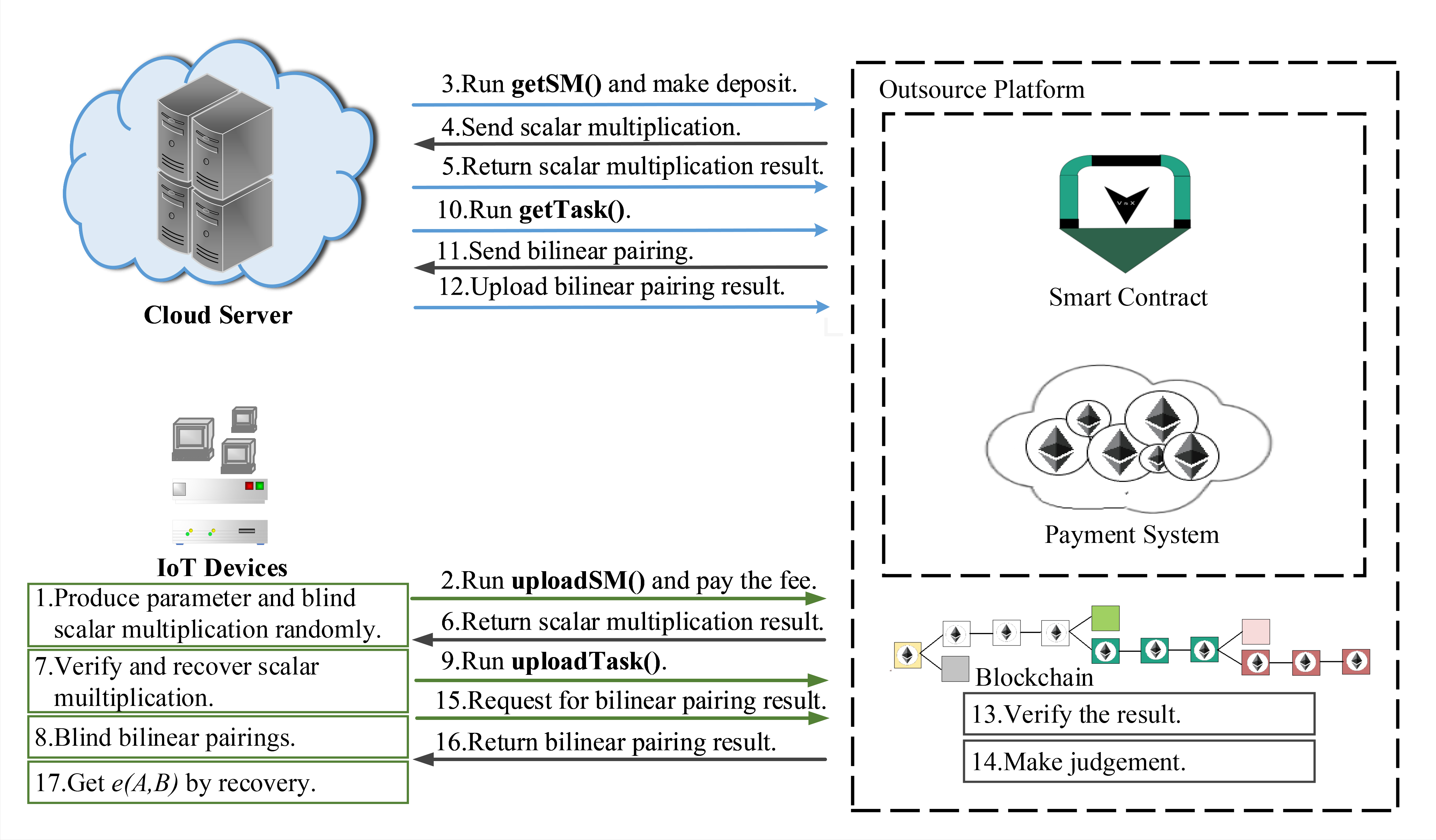}
	  	\caption{Blockchain-Based Fair Payment Scheme}
	\label{blockchain}
 \end{figure*}

    In this section, we apply the blockchain technology to enable fair payment for the secure outsourcing process. Figure~\ref{blockchain} shows the workflow of the blockchain-based fair payment system. As we can see, the client first generates some random parameters and obscure the inputs. Then the client runs the $uploadSM()$ smart contract function to upload the computations task onto the blockchain-based platform. Meanwhile, $T$ pays a service fee to the smart contract. When the cloud server $S$ wants to take a task from the platform, he runs the $getSM()$ smart contract function and gets the inputs of scalar multiplication. Meanwhile, $S$ needs to make a deposit to the payment system. Then $S$ conducts the scalar multiplication and returns the result. $T$ verifies the correctness of the scalar multiplication. Then $T$ obscures the inputs of bilinear pairing and calls $uploadTask()$ function to upload the task to the blockchain. Next, $S$ runs $getTask()$ function to get the bilinear paring task. $S$ conducts the bilinear paring operation and returns the result to the blockchain. The blockchain verifies the correctness of the calculation result and makes the judgment accordingly. If the result passes the verification, the smart contract will pay the service fee and return the deposit to the cloud server. If the result failed to pass, the smart contract would send the deposit and the service fee back to the client. 

    \section{Comparison}\label{com}

 \begin{figure}[t]
    	\centering
    	\includegraphics[height=6.2cm,width=9.0cm]{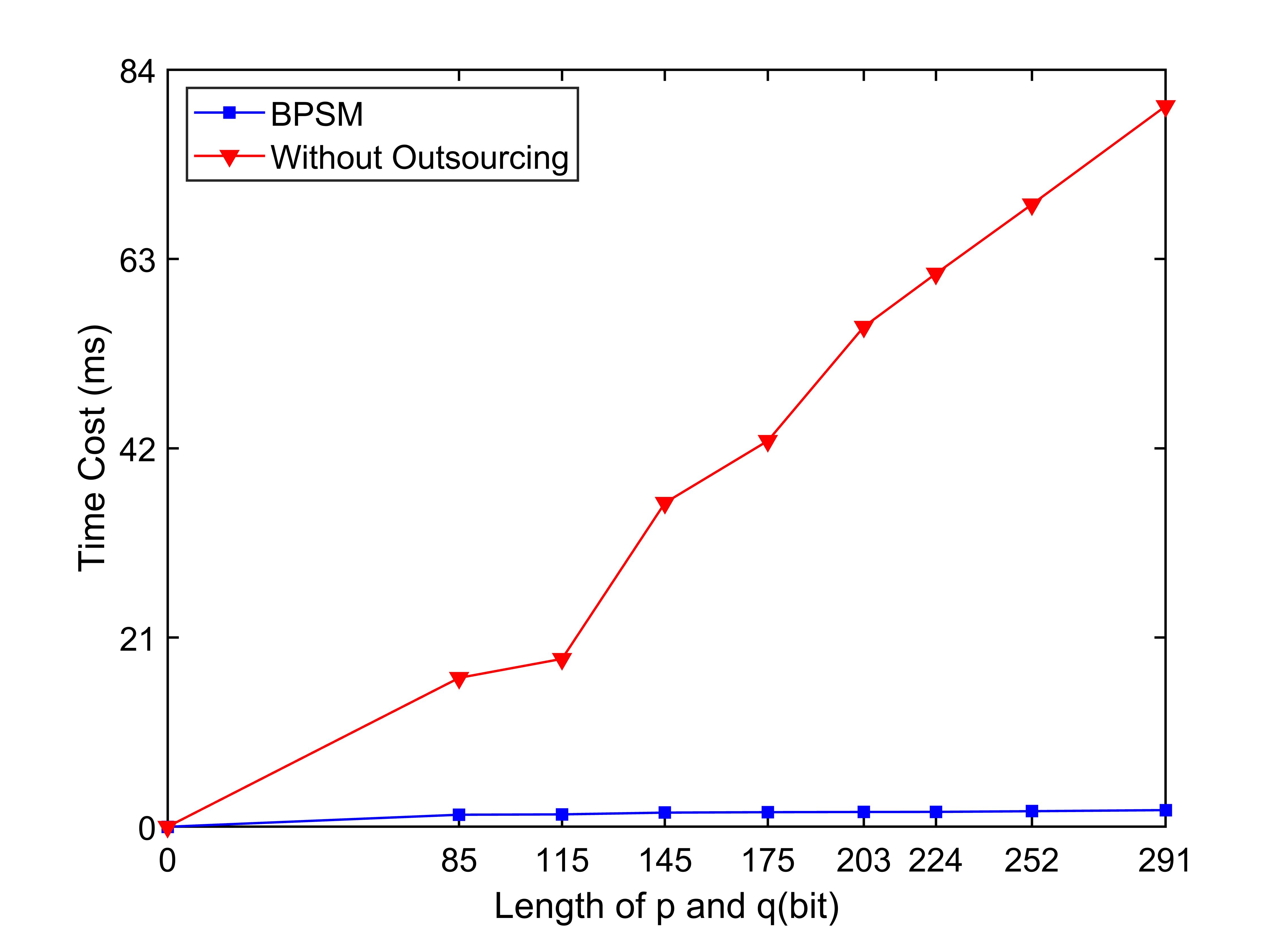}
    	\caption{Time Cost Comparison Between Client and Local Compute}
    	\label{img_cmpClientLocal}
    \end{figure}
    \begin{figure}[t]
    	\centering
    	\includegraphics[height=6.2cm,width=9.0cm]{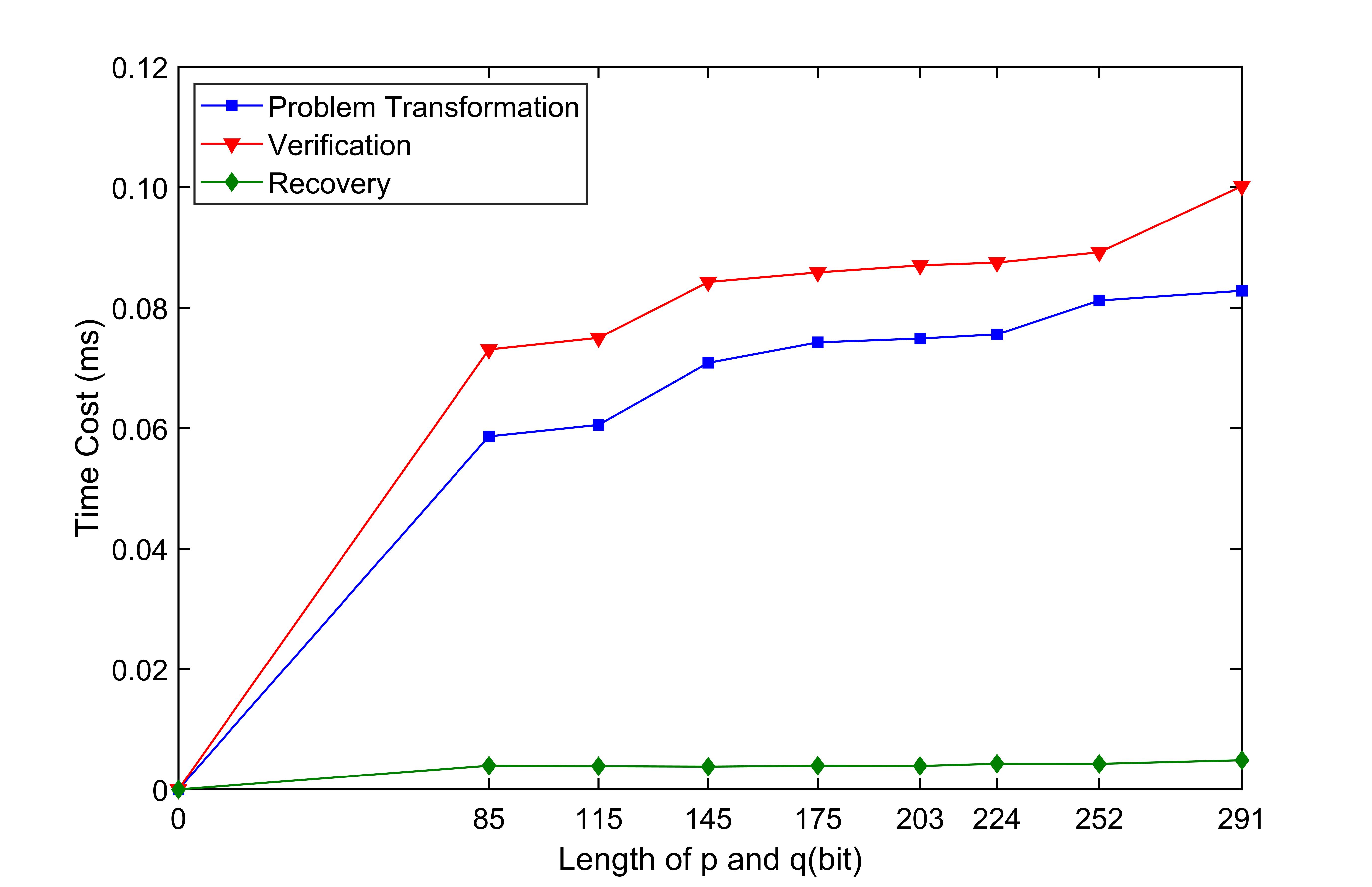}
    	\caption{Problem Transformation, Verification, Recovery Time of BPSM}
    	\label{img_cmpInner}
    \end{figure}
	\begin{table}
		    	\centering
		    	\caption{Comparison of Pre-computations}
		    	\begin{tabular}{cccc}
		    		\hline
		    		&BPSM& Alg\cite{chen2015efficient}& Alg\cite{tian2015secure}\\
		    		\hline
		    		$PA$& -& 0& $5(k+h-3)$\\
		    		$ME$& -& 0& 2\\
		    		$SM$& -& 9& 3\\
		    		$PE$& -& 3& 0\\
		    		$M$& -& 0& 0\\
		    		\hline
		    	\end{tabular}
		    	\label{table:PreCmp}
	\end{table}
	\begin{table}
		    	\centering
		    	\caption{Comparison of Client's Workload}
		    	\begin{tabular}{cccc}
		    		\hline
		    		&BPSM& Alg\cite{chen2015efficient}& Alg\cite{tian2015secure}\\
		    		\hline
		    		$PA$& 0& 5& 4\\
		    		$ME$& 1& 0& 0\\
		    		$SM$& 0& 0& 0\\
		    		$PE$& 0& 0& 0\\
		    		$M$& 2& 4& 3\\
		    		\hline
		    	\end{tabular}
		    	\label{table:ClientCmp}
	\end{table}
    \begin{table*}
    	\centering
    	\caption{Precomputation Storage of Alg\cite{chen2015efficient}}
    	\begin{tabular}{ccccc}
    		\hline
    		Elliptic Curves &10K& 100K& 1M& 10M\\
    		\hline
    		$d11499-85-82$&  4980KB& 48MB& 486MB& 4.75GB\\
    		$d277699-175-167$&  10MB& 100MB& 1001MB& 9.78GB\\
    		$d496659-224-224$& 13MB& 128MB& 1.25GB& 12.52GB\\
    		$d1003-291-247$&  17MB& 167MB& 1.63GB& 16.26GB\\
    		\hline
    	\end{tabular}
    	\label{table:StorageChen}
    \end{table*}
    \begin{table*}
    	\centering
    	\caption{Precomputation Storage of Alg\cite{tian2015secure}}
    	\begin{tabular}{ccccc}
    		\hline
    		Elliptic Curves &10K& 100K& 1M& 10M\\
    		\hline
    		$d11499-85-82$&  2905KB& 28MB& 284MB& 2.77GB\\
    		$d277699-175-167$&  5981KB& 58MB& 584MB& 5.70GB\\
    		$d496659-224-224$& 7656KB& 75MB& 748MB& 7.30GB\\
    		$d1003-291-247$&  9946KB& 97MB& 971MB& 9.49GB\\
    		\hline
    	\end{tabular}
    	\label{table:StorageTian}
    \end{table*}    
    
    	\subsection{Numeric Analysis}
		   	In this section, we compare our algorithm \textbf{BPSM} with algorithms \cite{chen2015efficient}, \cite{dong2018efficient} and \cite{tian2015secure}. In the following tables, we denote point addition in $\mathbb{G}_1$(or $\mathbb{G}_2$) as $PA$, modular exponentiation as $ME$, scalar multiplication as $SM$, pairing evaluation as $PE$, point multiplication in $\mathbb{G}_T$ as $M$. Table \ref{table:PreCmp} presents the comparison of pre-computations. Table \ref{table:ClientCmp} compares the client's computation. Table \ref{table:StorageChen} and Table \ref{table:StorageTian} show the storage space required to store the pre-computation results.

As we can observe from Table \ref{table:PreCmp}, the client in \textbf{BPSM} does not need to perform any expensive operations since there is no pre-computation in our algorithm \textbf{BPSM}. Alg\cite{chen2015efficient} requires 9 $SM$ and 3 $PE$. Alg\cite{tian2015secure} requires $5(k + h - 3)$ $PA$, where the value $k$ is the size of a set $S_1$. We can learn from \cite{nguyen2001distribution} that $k$ is around 20 and the value $h$ is less than 10. Thus, Alg~\cite{tian2015secure} requires about 135 $PA$, 2 $ME$ and 3 $SM$. Table \ref{table:ClientCmp} shows that the client only needs to conduct one modular exponentiation and two-point multiplications. And the power of modular exponentiation is less than the security parameter $m$. The computation conducted by the client in our algorithm \textbf{BPSM} is less than that in the other algorithms, which include the time-consuming point addition. Thus, our algorithm \textbf{BPSM} is more efficient on the client side than the other algorithms.

    	Table \ref{table:StorageChen} and Table \ref{table:StorageTian} show the required storage space for pre-computation results on the client side with different elliptic curves and different times of calculation. Note that the labels of elliptic curves are from the PBC library~\cite{lynnpbc} we used to implement the algorithms. As we can observe from Table \ref{table:StorageChen}, the pre-computation will occupy a large storage size of the client when conduct algorithm~\cite{chen2015efficient} many times. For example, with the $d1003-291-247$ elliptic curve, when the client conducts $10M$ times of pre-computations, it will require 16.26GB storage space, which is too much for resource-constrained IoT devices. Although the algorithm in Table \ref{table:StorageTian} has been improved on the basis of algorithm~\cite{chen2015efficient}, it still requires a lot of storage space. In contrast, Our algorithm does not require any pre-computation, which is more suitable to be applied on IoT devices.
		    \begin{figure}[t]
		       	\centering
		       	\includegraphics[height=6.2cm,width=9.0cm]{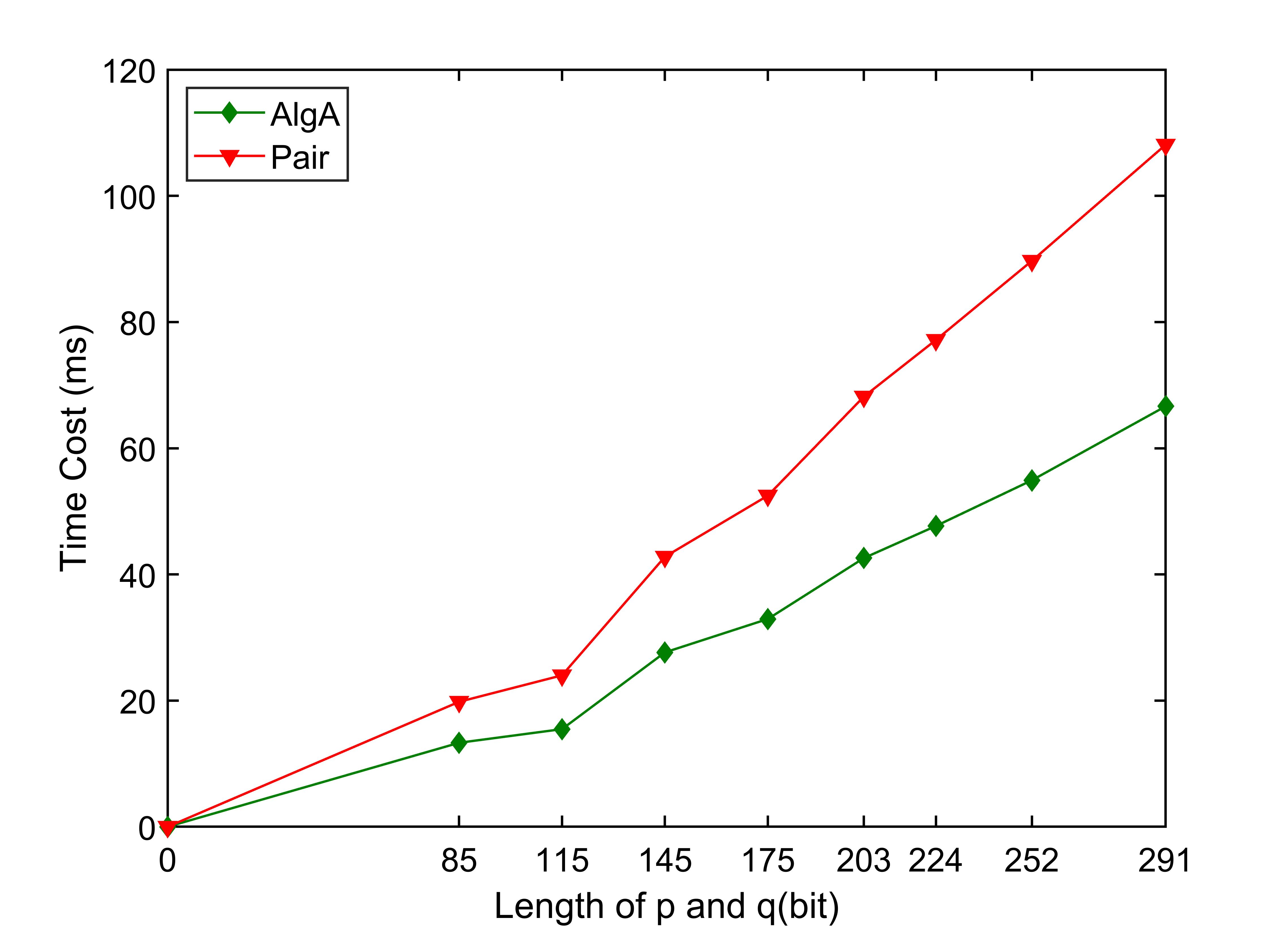}
		       	\caption{Pre-Computation Time Comparison Between Pair\cite{chen2015efficient} and AlgA\cite{tian2015secure} in IoT Device}
		       	\label{img_PApreTime}
		    \end{figure}
        \begin{figure}[t]
            \centering
            \includegraphics[height=6.2cm,width=9.0cm]{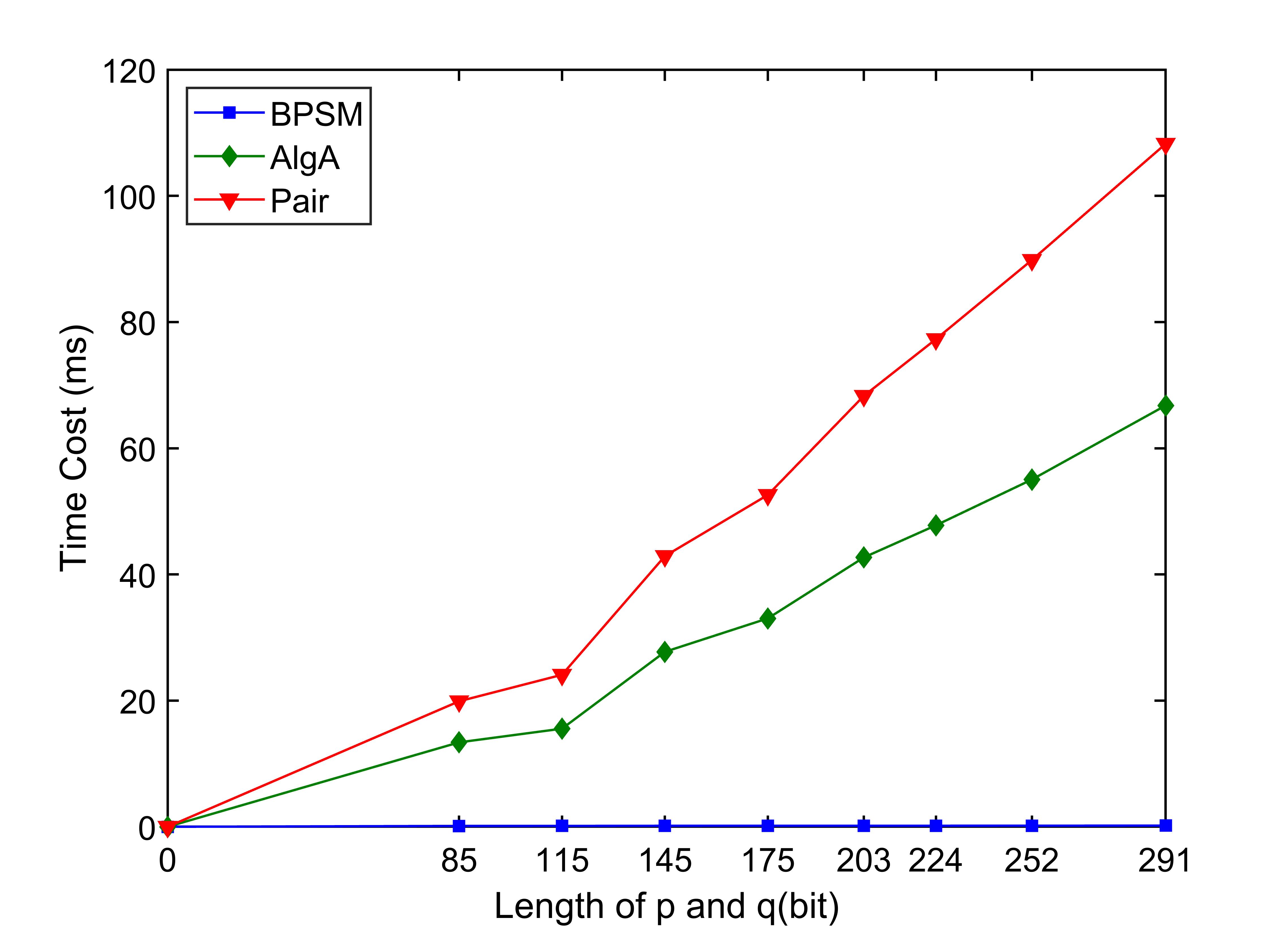}
            \caption{Time Cost of BPSM, Pair\cite{chen2015efficient} and AlgA~\cite{tian2015secure} in Client Side}
            \label{img_clientPreCmpBPA}
        \end{figure}
	   	\subsection{Performance Evaluation}
			\indent We implement our algorithms to demonstrate the practical efficiency. All data is the average value obtained by experimenting 500 rounds. In our experiment, the computations of the client and the cloud servers are conducted on the computer with Xeon E5-2620 processor running @2.4GHz with 1024MB RAM. The operating system is Ubuntu 17.10 x64. The program is developed in C++ with two open source libraries: GMP library and PBC library. We do not consider the communication consumption between the client and the cloud servers in our experiment. Our goal is to test the computation efficiency of our algorithm.
			\\
			\indent In Fig. \ref{img_cmpClientLocal}, we show the time cost on the client in our algorithm and the time required by computing the bilinear pairing locally. It is obvious that the time spent by the client in our algorithm is far less than that of computing the bilinear pairing locally. Fig. \ref{img_cmpInner} shows the time costs of the client in the three phases in our algorithm including blinding, verification and recovery. As shown in Fig. \ref{img_PApreTime}, algorithm \cite{chen2015efficient} and algorithm \cite{tian2015secure} both need to conduct time-consuming pre-computation. In IoT system, the pre-computation cannot be applied to IoT devices due to the insufficient storage resources of IoT devices. So the pre-computation of algorithm \cite{chen2015efficient} and algorithm \cite{tian2015secure} should be calculated by the IoT device in real time. In Fig. \ref{img_clientPreCmpBPA}, we compare the time costs of client among algorithm \cite{chen2015efficient}, algorithm~\cite{tian2015secure} and our algorithm in IoT system. We can see that the time cost of the client in our algorithm is significantly less than that in the other two algorithms. So the efficiency in client side of our algorithm is much higher than that of algorithm \cite{chen2015efficient} and algorithm \cite{tian2015secure} in IoT system.
			
	\section{Related Works}\label{rel}
		\indent Researchers have spent numerous efforts studying how to securely outsource varieties of computations so that resource-constrained devices can reduce the local workload. Gentry {\em et al.}~\cite{gentry2009fully} proposed a fully homomorphic encryption (FHE) algorithm so that users can achieve the goal of secure outsourcing with the help of an ideal lattice. Then, multiple general structures for secure outsourcing computation like \cite{chung2010improved, canard2013toward, brakerski2014leveled} have been proposed to realize various functions of computation, which are usually based on FHE algorithm. But these algorithms are usually inefficient as a result of their unpractical feature. Also, there are researchers who focus on solving specific problems. This kind of algorithm is usually much more efficient. For instance, Wei {\em et al.}~\cite{wei2019lightweight} first proposed a signature based on identity to achieve unforgeability against selected message attacks without random predictions. They designed two outsourcing algorithms for exponential operations, which is both secure and efficient, and it can reduce the computation cost of clients.

		\indent In cryptographic algorithm based on discrete logarithm, modular exponentiation is one of research hotspots due to its wide use. But it consumes time due to the large scale of the modular. Therefore, many studies have taken outsourcing modular exponentiation securely into consideration. Hohenberger {\em et al.}~\cite{hohenberger2005securely} presented the first solution. Then Chen {\em et al.}~\cite{chen2014new} proposed a more efficient algorithm in two untrusted program model, they also designed the first algorithm aimed at outsoucing the calculation of simultaneous modular exponentiations. Zhou {\em et al.}~\cite{zhou2017expsos} solved the problem of how to outsource exponentiation operations securely in one single untrusted program model. Their approcah provides a secure verification scheme in which the checkability is about 1. Ren {\em et al.}~\cite{ren2018efficient} proposed two algorithms about how to outsource modular exponentiation, which can detect malicious behaviors of the server. Liu {\em et al.}~\cite{liu2020verifiable} designed an innovative outsourcing solution about shareable functions, which is for modular exponentiation. It is secure even if there are adaptive adversaries. There are also some research works on outsourcing applications of modular exponentiation. For example, Zhang {\em et al.}~\cite{zhang2020efficient} designed a secure outsourcing algorithm for RSA Decryption, in which modular exponentiation is involved.

		\indent In many fields, such as image processing and machine learning, matrix operations are fundamental operations. Atallah {\em et al.}~\cite{atallah2010securely} first proposed a scheme of outsourcing matrix multiplication, which can be proved to be safe under the new calculation assumptions related to secret sharing. Zhang {\em et al.}~\cite{zhang2014efficient} proposed algorithms designed to outsource matrix multiplications, which is verifiable in both malicious and rational adversary model. Zhang {\em et al.}~\cite{zhang2017new} also focused on this subject and presented another public verifiable algorithm for matrix multiplication. Using random matrix blinding the original matrix, Mohassel {\em et al.}~\cite{mohassel2011efficient} designed for matrix inversion, while Lei {\em et al.}~\cite{lei2013outsourcing} proposed a scheme by using the matrix transformation technique. As for Xiao {\em et al.}~\cite{xiao2019novel}, they designed a neural network for the inversion of the time matrix solving in the complex field.
		
		\indent In mathematics, the definition of a linear system means it contains two or more linear equations with the same variable. Wang {\em et al.}~\cite{wang2013harnessing} first presented a secure outsourcing proposal using an iterative method for solving large linear equations. And Chen {\em et al.}~\cite{chen2014privacy} proposed a new protocol using special linear transformations. Different from the previous protocols, there is no homomorphic encryption and interaction between the client and the cloud.
	
		\indent Outsourcing bilinear pairing securely is also a hotspot because of the wide usage of bilinear pairing. Chevallier-Mames {\em et al.}~\cite{chevallier2010secure} designed a scheme for bilinear pairing. Devices which has constrained resource can detect malicious behaviors of servers. But many expensive operations still need to be executed by clients, which are time-consuming. Chen {\em et al.}~\cite{chen2015efficient} first considered algorithms for bilinear pairing by using precomputation. Tian {\em et al.}~\cite{tian2015secure} proposed a more efficient project-based on~\cite{chen2015efficient} under the same assumption with the same checkability. By introducing the pre-computation, Dong {\em et al.}~\cite{dong2018efficient} presented two sufficient secure schemes on the basis of a single untrusted server. But these pre-computation results depend on the large storage space of the client. Hence it's difficult for clients with limited computation and storage resources to realize all these algorithms. Lin {\em et al.}~\cite{lin2020blockchain} focused on it and proposed a novel blockchain-based system designed to efficiently solve the problem.

   	\section{Conclusion}\label{con}
   	In this paper, we explore how to delegate the bilinear pairing to untrustworthy cloud servers in a secure, fair, and efficient way. Existing algorithms cannot fit the IoT scenarios since they require extra storage space for the client. Our proposed algorithm solves this problem by coming up with new strategies to obscure the inputs. To ensure the fairness of payment, we construct a fair payment framework on the Ethereum blockchain. Our developed smart contract ensures that the cloud server gets paid only when he correctly performed the bilinear pairing for the client. We also evaluate our proposed algorithm through theoretical analysis and experiments in which privacy, fairness, and efficiency are justified.

   \bibliographystyle{plain}
\bibliography{Ref_IoT_BPSM}

\begin{thebibliography}{10}

\bibitem{armbrust2010view}
Michael Armbrust, Armando Fox, Rean Griffith, Anthony~D Joseph, Randy Katz,
  Andy Konwinski, Gunho Lee, David Patterson, Ariel Rabkin, Ion Stoica, et~al.
\newblock A view of cloud computing.
\newblock {\em Communications of the ACM}, 53(4):50--58, 2010.

\bibitem{atallah2010securely}
Mikhail~J Atallah and Keith~B Frikken.
\newblock Securely outsourcing linear algebra computations.
\newblock In {\em Proceedings of the 5th ACM Symposium on Information, Computer
  and Communications Security}, pages 48--59. ACM, 2010.

\bibitem{brakerski2014leveled}
Zvika Brakerski, Craig Gentry, and Vinod Vaikuntanathan.
\newblock (leveled) fully homomorphic encryption without bootstrapping.
\newblock {\em ACM Transactions on Computation Theory (TOCT)}, 6(3):13, 2014.

\bibitem{canard2013toward}
S{\'e}bastien Canard, Iwen Coisel, Julien Devigne, C{\'e}cilia Gallais, Thomas
  Peters, and Olivier Sanders.
\newblock Toward generic method for server-aided cryptography.
\newblock In {\em International Conference on Information and Communications
  Security}, pages 373--392. Springer, 2013.

\bibitem{chen2014privacy}
Fei Chen, Tao Xiang, and Yuanyuan Yang.
\newblock Privacy-preserving and verifiable protocols for scientific
  computation outsourcing to the cloud.
\newblock {\em Journal of Parallel and Distributed Computing},
  74(3):2141--2151, 2014.

\bibitem{chen2014new}
Xiaofeng Chen, Jin Li, Jianfeng Ma, Qiang Tang, and Wenjing Lou.
\newblock New algorithms for secure outsourcing of modular exponentiations.
\newblock {\em IEEE Transactions on Parallel and Distributed Systems},
  25(9):2386--2396, 2014.

\bibitem{chen2015efficient}
Xiaofeng Chen, Willy Susilo, Jin Li, Duncan~S Wong, Jianfeng Ma, Shaohua Tang,
  and Qiang Tang.
\newblock Efficient algorithms for secure outsourcing of bilinear pairings.
\newblock {\em Theoretical Computer Science}, 562:112--121, 2015.

\bibitem{chevallier2010secure}
Beno{\^\i}t Chevallier-Mames, Jean-S{\'e}bastien Coron, Noel McCullagh, David
  Naccache, and Michael Scott.
\newblock Secure delegation of elliptic-curve pairing.
\newblock In {\em International Conference on Smart Card Research and Advanced
  Applications}, pages 24--35. Springer, 2010.

\bibitem{chung2010improved}
Kai-Min Chung, Yael Kalai, and Salil Vadhan.
\newblock Improved delegation of computation using fully homomorphic
  encryption.
\newblock In {\em Annual Cryptology Conference}, pages 483--501. Springer,
  2010.

\bibitem{dong2018efficient}
Min Dong and Yanli Ren.
\newblock Efficient and secure outsourcing of bilinear pairings with single
  server.
\newblock {\em Science China Information Sciences}, 61(3):039104, 2018.

\bibitem{gentry2009fully}
Craig Gentry and Dan Boneh.
\newblock {\em A fully homomorphic encryption scheme}, volume~20.
\newblock Stanford University Stanford, 2009.

\bibitem{hohenberger2005securely}
Susan Hohenberger and Anna Lysyanskaya.
\newblock How to securely outsource cryptographic computations.
\newblock In {\em Theory of Cryptography Conference}, pages 264--282. Springer,
  2005.

\bibitem{lei2013outsourcing}
Xinyu Lei, Xiaofeng Liao, Tingwen Huang, Huaqing Li, and Chunqiang Hu.
\newblock Outsourcing large matrix inversion computation to a public cloud.
\newblock {\em IEEE Transactions on cloud computing}, 1(1):1--1, 2013.

\bibitem{lin2020blockchain}
Chao Lin, Debiao He, Xinyi Huang, Xiang Xie, and Kim-Kwang~Raymond Choo.
\newblock Blockchain-based system for secure outsourcing of bilinear pairings.
\newblock {\em Information Sciences}, 527:590--601, 2020.

\bibitem{lin2017survey}
Jie Lin, Wei Yu, Nan Zhang, Xinyu Yang, Hanlin Zhang, and Wei Zhao.
\newblock A survey on internet of things: Architecture, enabling technologies,
  security and privacy, and applications.
\newblock {\em IEEE Internet of Things Journal}, 4(5):1125--1142, 2017.

\bibitem{liu2020verifiable}
Muhua Liu, Ying Wu, Rui Xue, and Rui Zhang.
\newblock Verifiable outsourcing computation for modular exponentiation from
  shareable functions.
\newblock {\em Cluster Computing}, 23(1):43--55, 2020.

\bibitem{lynnpbc}
Ben Lynn.
\newblock Pbc library, 2006.
\newblock {\em URL http://crypto. stanford. edu/pbc}.

\bibitem{mohassel2011efficient}
Payman Mohassel.
\newblock Efficient and secure delegation of linear algebra.
\newblock {\em IACR Cryptology ePrint Archive}, 2011:605, 2011.

\bibitem{nguyen2001distribution}
Phong~Q Nguyen, Igor~E Shparlinski, and Jacques Stern.
\newblock Distribution of modular sums and the security of the server aided
  exponentiation.
\newblock In {\em Cryptography and Computational Number Theory}, pages
  331--342. Springer, 2001.

\bibitem{ren2012security}
Kui Ren, Cong Wang, and Qian Wang.
\newblock Security challenges for the public cloud.
\newblock {\em IEEE Internet Computing}, 16(1):69--73, 2012.

\bibitem{ren2018efficient}
Yanli Ren, Min Dong, Zhenxing Qian, Xinpeng Zhang, and Guorui Feng.
\newblock Efficient algorithm for secure outsourcing of modular exponentiation
  with single server.
\newblock {\em IEEE Transactions on Cloud Computing}, 2018.

\bibitem{tian2015secure}
Haibo Tian, Fangguo Zhang, and Kun Ren.
\newblock Secure bilinear pairing outsourcing made more efficient and flexible.
\newblock In {\em Proceedings of the 10th ACM Symposium on Information,
  Computer and Communications Security}, pages 417--426. ACM, 2015.

\bibitem{TYZ19}
Le~Tong, Jia Yu, and Hanlin Zhang.
\newblock Secure outsourcing algorithm for bilinear pairings without
  pre-computation.
\newblock In {\em 2019 IEEE Conference on Dependable and Secure Computing
  (DSC)}, pages 1--7. IEEE, 2019.

\bibitem{vaquero2008break}
Luis~M Vaquero, Luis Rodero-Merino, Juan Caceres, and Maik Lindner.
\newblock A break in the clouds: towards a cloud definition.
\newblock {\em ACM SIGCOMM Computer Communication Review}, 39(1):50--55, 2008.

\bibitem{wang2013harnessing}
Cong Wang, Kui Ren, Jia Wang, and Karthik Mahendra.
\newblock Harnessing the cloud for securely outsourcing large-scale systems of
  linear equations.
\newblock {\em IEEE Transactions on Parallel and Distributed Systems},
  24:1172--1181, 2013.

\bibitem{wei2019lightweight}
Zhijun Wei, Jing Li, Xianmin Wang, and Chong-Zhi Gao.
\newblock A lightweight privacy-preserving protocol for vanets based on secure
  outsourcing computing.
\newblock {\em IEEE Access}, 7:62785--62793, 2019.

\bibitem{xiao2019novel}
Lin Xiao, Yongsheng Zhang, Kenli Li, Bolin Liao, and Zhiguo Tan.
\newblock A novel recurrent neural network and its finite-time solution to
  time-varying complex matrix inversion.
\newblock {\em Neurocomputing}, 331:483--492, 2019.

\bibitem{yu2016enabling}
Jia Yu, Kui Ren, and Cong Wang.
\newblock Enabling cloud storage auditing with verifiable outsourcing of key
  updates.
\newblock {\em IEEE Transactions on Information Forensics and Security},
  11(6):1362--1375, 2016.

\bibitem{yu2015enabling}
Jia Yu, Kui Ren, Cong Wang, and Vijay Varadharajan.
\newblock Enabling cloud storage auditing with key-exposure resistance.
\newblock {\em IEEE Transactions on Information forensics and security},
  10(6):1167--1179, 2015.

\bibitem{yu2017strong}
Jia Yu and Huaqun Wang.
\newblock Strong key-exposure resilient auditing for secure cloud storage.
\newblock {\em IEEE Transactions on Information Forensics and Security},
  12(8):1931--1940, 2017.

\bibitem{yu2017survey}
Wei Yu, Fan Liang, Xiaofei He, William~Grant Hatcher, Chao Lu, Jie Lin, and
  Xinyu Yang.
\newblock A survey on the edge computing for the internet of things.
\newblock {\em IEEE Access}, 6:6900--6919, 2017.

\bibitem{zhang2020efficient}
Hanlin Zhang, Jia Yu, Chengliang Tian, Le~Tong, Jie Lin, Linqiang Ge, and
  Huaqun Wang.
\newblock Efficient and secure outsourcing scheme for rsa decryption in
  internet of things.
\newblock {\em IEEE Internet of Things Journal}, 2020.

\bibitem{zhang2017new}
Xiaoyu Zhang, Tao Jiang, Kuan-Ching Li, Aniello Castiglione, and Xiaofeng Chen.
\newblock New publicly verifiable computation for batch matrix multiplication.
\newblock {\em Information Sciences}, 2017.

\bibitem{zhang2014efficient}
Yihua Zhang and Marina Blanton.
\newblock Efficient secure and verifiable outsourcing of matrix
  multiplications.
\newblock In {\em International Conference on Information Security}, pages
  158--178. Springer, 2014.

\bibitem{zhou2017expsos}
Kai Zhou, MH~Afifi, and Jian Ren.
\newblock Expsos: Secure and verifiable outsourcing of exponentiation
  operations for mobile cloud computing.
\newblock {\em IEEE Transactions on Information Forensics and Security},
  12(11):2518--2531, 2017.

\end{thebibliography}
\end{document}